\documentclass{IEEEtran4PSCC}

\ifCLASSINFOpdf
   \usepackage[pdftex]{graphicx}
\else
   \usepackage[dvips]{graphicx}
\fi

\usepackage[cmex10]{amsmath}

\usepackage{amsfonts}
\usepackage{color}

\usepackage{nomencl}
\usepackage{float}
\makenomenclature
\usepackage{multirow}

\makeatletter
\let\old@ps@headings\ps@headings
\let\old@ps@IEEEtitlepagestyle\ps@IEEEtitlepagestyle
\def\psccfooter#1{%
    \def\ps@headings{%
        \old@ps@headings%
        \def\@oddfoot{\strut\hfill#1\hfill\strut}%
        \def\@evenfoot{\strut\hfill#1\hfill\strut}%
    }%
    \def\ps@IEEEtitlepagestyle{%
        \old@ps@IEEEtitlepagestyle%
        \def\@oddfoot{\strut\hfill#1\hfill\strut}%
        \def\@evenfoot{\strut\hfill#1\hfill\strut}%
    }%
    \ps@headings%
}
\makeatother

	\psccfooter{%
        \parbox{\textwidth}{\hrulefill \\ \small{22nd Power Systems Computation Conference} \hfill \begin{minipage}{0.2\textwidth}\centering \vspace*{4pt} \includegraphics[scale=0.06]{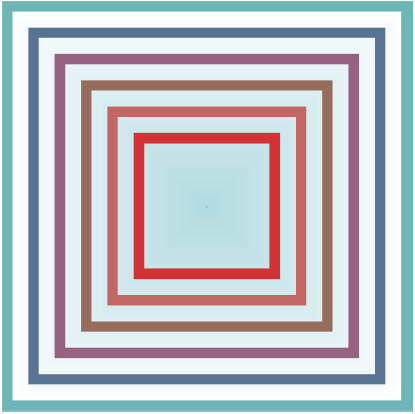}\\\small{PSCC 2022} \end{minipage} \hfill \small{Porto, Portugal --- June 27 -- July 1, 2022}}%
}

\begin{document}
%

\title{Efficient Representations of Radiality Constraints in \\ Optimization of Islanding and De-Energization \\ in Distribution Grids}

\author{
\IEEEauthorblockN{Joe Gorka and Line Roald}
\IEEEauthorblockA{
University of Wisconsin,
Madison, USA, 
\{jgorka, roald\}@wisc.edu}
}


\maketitle

\begin{abstract}
Optimization of power distribution system topology is complicated by the requirement that the system be operated in a radial configuration.  
In this paper, we discuss existing methods for enforcing radiality constraints and introduce two new formulations that enable optimization over partially energized or islanded network topologies. The first builds on methods that use so-called parent-child constraints, but enforces those constraints on an abstracted network which enables an equivalent formulation with significantly less variables and constraints.
The second formulation builds on existing approaches which directly generate constraints disallowing loops, and through an iterative approach seeks to limit the number of these constraints which must be enforced. 
These two new radiality constraint methods as well as two existing approaches are incorporated in otherwise identical OPF formulations which model an optimal power shut-off problem, where 
the utility of partially energized network configurations is harnessed to mitigate wildfire risk. 
Through tests on a medium-sized distribution feeder, we show that the two proposed approaches are significantly faster than existing methods and reliably obtain optimal solutions.


\end{abstract}

\begin{IEEEkeywords}
Distribution grid optimization, radiality constraints, de-energization
\end{IEEEkeywords}

\thanksto{\noindent Submitted to the 22nd Power Systems Computation Conference (PSCC 2022).}

\section{Introduction}
\subsection{Motivation}
Greater penetration of variable renewable energy resources 
and more frequent incidence of severe weather events due to global warming is increasing the need for flexible power distribution systems. Allowing changes in distribution network topology is one useful method for introducing such flexibility. Topology modification/optimization is typically used to restore  power to customers following a fault \cite{arif2018dynamic,poudel2020advanced,  chen2015resilient}, and has been proposed for load balancing and loss minimization \cite{liu1989loss, sarfi1994survey, jabr2012minimum,Baran, chiang1990optimal1, chiang1990optimal2}. 
In contrast to transmission grids, which are typically operated as a meshed system (i.e. with loops), distribution grids are generally operated in radial configurations (i.e. with a tree-like topology) in order to simplify fault isolation and protection. While distribution grids contain switches to enable topology changes, any change must result in a radial configuration. When optimizing distribution system topology, it is therefore necessary to include \textit{radiality constraints} to preclude the formation of any loops. Because of this, radiality constraints are a key ingredient in a wide variety of problems.

The use of switching to improve operations, improve fault isolation and reduce restoration times has received increased attention in recent years \cite{arif2018dynamic, poudel2020advanced, dirkman2014pe}. 
An emerging problem of interest is wildfire risk mitigation, where utilities implement public safety power shut-offs to prevent wildfire ignitions. 
When balancing risk minimization and load served in the context of wildfire prevention,
flexible power distribution grids are preferable, as they allow for more targeted power shut-offs \cite{PGE-2020safetyplan}.
However, flexibility-enhancing
grid features such as distributed generation and switching capabilities complicate the task of maintaining radial configurations. 
To address this problem, this paper proposes and benchmarks different formulations of radiality constraints, 
which are capable of ensuring radiality even in complex network configurations such as partially-energized states or islanded microgrids. 

\subsection{Relevant Literature on Radiality in Distribution Systems}
Research on distribution system reconfiguration initially focused on heuristics, before eventually employing meta-heuristics and formal optimization techniques. An excellent description of important heuristic methods can be found in \cite{jabr2012minimum}, summarized as follows:
The method in \cite{Merlin} starts from a configuration where all switches are closed, then successively open switches based on a proxy for loss minimization until radiality is achieved. The method in \cite{Civanlar} instead starts from a radial configuration and flips the statuses of a chosen sequence of switch pairs (one closed and one open) to minimize losses while maintaining radiality. 
Methods based on \cite{Civanlar} made modifications to switching operation criteria such as the measure of network losses \cite{Baran,Gomes,Schmidt,Raju}, or switch locations \cite{Goswami}.

While heuristic approaches are quite fast, methods based on mathematical optimization have the benefit of guaranteeing optimality of the solution. 
There are two main types of radiality constraint formulations in the literature, loop-based formulations and parent-child formulations. 
Loop-based formulations, which enforce radiality by explicitly requiring that each potential loop contain an open switch, have been used in \cite{arif_thesis, barnes2019resilient}. However, while loop-based formulations are conceptually simple, they introduce one constraint per loop and the process of producing and enforcing these constraints can represent a significant computational burden. 
Since the direction of power flow in distribution system cannot be determined a priori, the problem of enumerating all potential loops is equivalent to finding all simple cycles in an undirected graph. Efficient methods have been proposed for the enumeration of cycles in directed graphs \cite{Tarjan} or restricted classes of cycles in undirected graphs \cite{Uno}, but the authors are not aware of any successful attempts to extend these to the enumeration of general simple cycles in undirected graphs. In \cite{Borghetti}, a clever method to solve this problem through the construction of a loop basis is proposed, 
but it remains unclear how 
this compares to immediately seeking to enumerate all cycles.

Another category of radiality constraint that should be acknowledged is those which exactly specify the number of active lines in the system, in general requiring that the number of active lines be equal to one less than the number of buses. Though this constraint alone was shown by \cite{Ahmadi2015_radiality_constraints} to be insufficient, methods such as that proposed in \cite{Lavorato2012_radiality_constraints} have added additional constraints to fix this issue and enable consideration of multiple substations/islanded operation of DERs. Despite this, such methods require that decisions on whether or not to island a portion of the grid are made a priori. Additionally, there is no support for operating the grid in partially-energized configurations.

In contrast to loop-based formulations and approaches which constrain global properties like the number of active lines, so-called parent-child formulations prevent loops through local inter-nodal relationships. 
In \cite{jabr2012minimum}, a radiality constraint formulation was introduced which made use of parent-child relationships between adjacent network nodes to shape the network into a spanning tree. The substation node, required to have no parent nodes, functioned as the root of the tree. 
While the constraints from \cite{jabr2012minimum} ensure radiality under many conditions, \cite{Ahmadi2015_radiality_constraints} showed that they no longer guarantee radiality if the network is split into several connected components.
Such disconnected topologies are often infeasible with respect to the power flow equations and are discarded, somewhat masking this deficiency \cite{Ahmadi}. However, there are cases in which such split configurations are in fact power flow feasible due to distributed generation or back-feeding from another substation, and/or are necessary to model partly de-energized networks due to, e.g., fault isolation \cite{poudel2020advanced} or wildfire-related public safety power shut-offs \cite{rhodes2020balancing}.
In \cite{arif2018dynamic} the radiality constraints presented in \cite{jabr2012minimum} were extended to ensure the radiality of networks in such configurations by introducing the concept of a virtual source and virtual flow. 


\subsection{Contributions}
A weakness of the parent-child approach to radiality constraints, as defined in \cite{jabr2012minimum}, is that it is necessary to associate two binary variables with each line in order to track the parent-child relationship between its endpoint nodes. In also representing the relationship of each node with the virtual source, the formulation in \cite{arif2018dynamic} requires an even greater number of binary variables. 
Furthermore, the loop-based approach suffers from the lack of efficient loop enumeration algorithms, limiting the scalability of the method.
Thus, the existing methods to represent radiality constraints have shown severe limits on scalability. This paper attempts to address this problem by proposing two new methods for enforcing radiality constraints, (i) an equivalent parent-child formulation based on an abstracted network representation and (ii) an iterative approach to identify only the most important loop constraints. 
We further introduce an optimization model to identify optimal power shut-offs in the context of wildfire risk into which the different radiality constraint formulations are integrated. This model allows us to test the efficacy of our formulations in the context of a problem that includes both de-energization and islanded operation. In our a case study, we compare the proposed radiality constraint formulations to existing approaches both in terms of the number of constraints and variables, and in terms of solution times. We find that the proposed formulations achieve speed-ups of more than 10-20 times on challenging instances. 



\section{Optimizing Islanding and De-Energization in Distribution Grid Operations}

The focus of the distribution grid optimization problem discussed in this paper is to optimize the system topology to (i) isolate parts of the system that need to be de-energized and (ii) form islanded microgrids while maintaining a radial grid topology. While similar problems arise in the context of fault isolation and post-fault restoration, we consider the problem of implementing public safety power shut-offs where parts of the grid is de-energized to reduce the risk of wildfire ignitions. 
Specifically, we are interested in understanding how to best balance wildfire risk mitigation (through de-energization) and the wish to continue to serve power to as many customers as possible. A related version of this problem, which we refer to as the optimal power shut-off problem, was discussed for transmission grids in \cite{rhodes2020balancing}. Here, we develop a similar, yet distinct version of this problem that allows us to investigate shut-offs in the distribution grid. 
This section first presents a proposed abstraction for distribution grids, before discussing the mathematical formulation of the optimal shut-off problem.


\subsection{Distribution Grid Abstraction}
\label{sec:abstraction}



We consider a power distribution network with with a set of nodes $\mathcal{N}$, a set of lines $\mathcal{L}$ and a set of lines with switches $\mathcal{S}\subset\mathcal{L}$. Nodes are indexed by single subscripts $n\in\mathcal{N}$, while lines are indexed by their number $l$, their from bus $i$ and their to bus $j$, i.e. $lij\in\mathcal{L}$. The indexing variable $l$ is employed to accommodate a situation where two nodes are connected by multiple switched lines.

\subsubsection{Definition of Load Blocks}
Because switches are not available at each line and each node, shutting down an individual line or node is typically not possible. 
As a result, if we want to de-energize a given line or node, we have to de-energize a larger part of the network. We refer to the parts that can be de-energized at the same time as \emph{load blocks}, 
i.e. parts of the network that are internally connected by non-switchable lines and connected to other load blocks through switchable lines. The load blocks can be identified by opening all of the switches in a power system and noting the islands that form, and each block has an internal topology that is immutable (i.e., cannot be changed). We denote the set of load blocks by $\mathcal{B}$, the set of nodes $i$ belonging in load block $m$ as $i\in\mathcal{N}_m$, and the mapping between nodes and load blocks through the vector $NB\in\mathbb{R}^{|\mathcal{N}|}$ which contains the assigned load block $b$ for each node $n$, i.e., $NB_{n}=b$.  


\subsubsection{Network abstraction}
In our proposed network abstraction, the load blocks $\mathcal{B}$ make up the nodes, while switched lines $\mathcal{S}$, which represent the connections between load blocks, make up the edges. 

The abstracted network is helpful when discussing both partial de-energization of the grid and constraints that enforce radial topologies. 
\begin{itemize}
    \item \emph{De-energization:} Since it is necessary to isolate energized and de-energized parts of the grid by a switch, the load blocks represent the smallest units for which we can make independent energization/de-energization decisions. 
    \item \emph{Radial operation:} For radial grid operation to be possible (regardless of switch status) each of these load blocks are internally radial, i.e., there are no loops within them. As a result, the radiality of the network as a whole can be assured by preventing loops in the abstracted network, i.e. loops made up of load blocks. 
\end{itemize}


\subsubsection{Switching and Energization Constraints in the Abstracted Network}

For each load block $m\in\mathcal{B}$ we define a binary energization variable $z_m$, where $z_m=1$ and $z_m=0$ indicates that the node is energized or de-energized, respectively. For the lines with switches, define the binary variable $s_{lmn}$, where $s_{lmn}=1$ denotes a closed switch and $s_{lmn}=0$ an open switch. 
These energization $z$ and switching $s$ govern the energization status and topology of our grid. 
A general requirement for the switching status is as follows: 
if two load blocks are connected by a closed switch, then they have the same energization status (i.e., they are either both energized or both de-energized). Stated mathematically, we have that any two load blocks $m$ and $n$ connected by a switched line $s_{lmn}$ must satisfy the following constraints,
\begin{subequations}
\label{eq:switching}
\begin{align}
    z_m  &\geq s_{lmn} + z_n -1~, \\
    z_n  &\geq s_{lmn} + z_m -1~.
\end{align}
\end{subequations}
If $s_{lmn} = 1$ (the switch is closed), then $z_m=z_n$. Otherwise if $s_{lmn} = 0$ (the switch is open), there is no constraint on the energization statuses of $m$ and $n$ ($z_m$ and $z_n$) with respect to each other.
It can be verified that these constraints ensure that de-energized load blocks are properly separated from the energized parts of the network, but they do not preclude the possibility of an islanded load block that remains energized through distributed generation (despite being disconnected from the substation).

\subsection{Optimal Islanding and De-Energization}
In the following, we describe the mathematical formulation for a distribution grid problem with islanding and de-energization. To make the formulation concrete, we discuss optimal power shut-off problem to reduce wildfire risk, but the formulation could easily be adapted to consider other objectives or problem variants. 

\subsubsection{Objective Function}
Our objective is to minimize total risk of the system, while maximizing the total load served to all customers. This is expressed as 
\begin{align} 
min \sum_{b \in \mathcal{B}} R_b z_b(1-\alpha)
- \sum_{b \in \mathcal{B}}D_b z_b \alpha
\label{eq:objective}
\end{align}
\begin{subequations}
\label{opf-formulation}
Here, the first term represent the wildfire risk of the load blocks. The wildfire risk associated with keeping load block $b$ energized is represented by $R_b$, which is the sum of the wildfire risk associated with all lines in load block $b$, i.e. $R_b=\sum_{l\in\mathcal{L}_b}R_l$ where $\mathcal{L}_b$ is the set of lines in load block $b$ and $R_l$ is their corresponding wildfire risk. 
Examples of how to define the risk values $R_l$ can be found in, e.g., \cite{rhodes2020balancing, taylor2021_wildfire_risk}.
If the load block is de-energized $z_b=0$, the wildfire risk is reduced to zero.
The second term of the objective function tracks how much total load is being served, with $D_b$ representing the sum of the loads at all nodes in load block $b$, i.e. $D_b=\sum_{k\in \mathcal{N}_b}p_{dk}$ where $p_{dk}$ is the active power demand on node k. If $z_b=0$, no load is served. 
The $\alpha$ is a trade-off parameter which determines the relative importance of each part of the objective. 
Choosing different values for $\alpha$ allows for adjustment between the extremes of total focus on risk minimization ($\alpha = 0$) or on the maximization of load served ($\alpha = 1$).

\subsubsection{Power Flow and Voltage Constraints}
The power flow and voltage at each bus are calculated using a linearized and lossless version of the DistFlow equations, referred to as LinDistFlow \cite{baran1989optimal}, but adapted to include energization variables for nodes and lines. 
Eqs. \eqref{eq:nodal_balance_p}, \eqref{eq:nodal_balance_q} represent the active and reactive power balance of each node $i\in\mathcal{N}$, with $z_b$ accounting for the energization status of the corresponding load block and $w_i$ representing the voltage magnitude squared. The variables $p_{lij}, q_{lij}$ represent the active and reactive power flows on line $lij$. The sets $G_i$, $D_i$, and $H_i$ represent the generators, loads, and shunt elements, respectively, which are connected at node $i$. Sums are performed over individual elements $k$ within these sets, necessitating the variables $p_{gk}$ and $q_{gk}$ for active and reactive power generated by generator $k$, $p_{dk}$ and $q_{dk}$ for active and reactive power demanded by load $k$, and finally $g_{hk}$ and $b_{hk}$ for the shunt conductance and susceptance of shunt element $k$. The set $L_i^+$ refers to the lines $lij \in \mathcal{L}$ for which node $i$ is the 'from' node, while $L_i^-$ refers to lines $lji \in \mathcal{L}$ for which node $i$ is the 'to' node.
\begin{align}
&\sum_{l\in L_i^+}{\!p_{lij}} - \!\! \sum_{l\in L_i^-}{\!p_{lji}} = \! z_m \!\left(\sum_{k\in G_i}{\!p_{gk}} - \!\! \sum_{k\in D_i}{\!p_{dk}} - \!\! \sum_{k\in H_i}{\!g_{hk}w_i}\!\right) \nonumber\\
&\quad\quad\quad\quad\quad\quad\quad\quad\quad\quad\quad\text{ with } m=NB_{i},~ \forall i\in\mathcal{N}\label{eq:nodal_balance_p}
\\
&\sum_{l\in L_i^+}{\!q_{lij}} - \!\! \sum_{l\in L_i^-}{\!q_{lji}} = \! z_m \!\left(\sum_{k\in G_i}{\!q_{gk}} - \!\! \sum_{k\in D_i}{\!q_{dk}} - \!\! \sum_{k\in H_i}{\!b_{hk}w_i}\!\right)  \nonumber\\
&\quad\quad\quad\quad\quad\quad\quad\quad\quad\quad\quad\text{ with } m=NB_{i},~ \forall i\in\mathcal{N}\label{eq:nodal_balance_q}
\end{align}
Eqs. \eqref{eq:ns_pf_1}-\eqref{eq:s_pf_2} represent the LinDistFlow voltage drop equations for switched lines \eqref{eq:ns_pf_1}, \eqref{eq:ns_pf_2} and non-switched lines \eqref{eq:s_pf_1}, \eqref{eq:s_pf_2}. While the standard LinDistFlow equations are linear equality constraints, we represent them equivalently as two linear inequality constraints to account for energization and switching status of the lines. 
\begin{align}
w_j &\leq w_i-2(r_lp_{lik} + x_lq_{lik})+M(1-z_m) \label{eq:ns_pf_1}\\
w_j &\geq w_i-2(r_lp_{lik} + x_lq_{lik})-M(1-z_m) \label{eq:ns_pf_2}\\
&\quad\quad\quad\quad\forall (lij)\in\mathcal{L},~\text{with } NB_i=NB_j=m \nonumber\\
w_j &\leq w_i-2(r_lp_{lik} + x_lq_{lik})+M(1-s_{lmn}) \label{eq:s_pf_1}\\
w_j &\geq w_i-2(r_lp_{lik} + x_lq_{lik})-M(1-s_{lmn}) \label{eq:s_pf_2} \\
&\quad\quad\quad\quad\forall lmn\in\mathcal{S},~\text{with } NB_i=m,~NB_j=n \nonumber
\end{align}
Eqs. \eqref{eq:ns_pf_1}, \eqref{eq:ns_pf_2} represent the case where the two nodes $i$ and $j$ belong to the same load block with a common the energization variable $z_m$. 
Eqs. \eqref{eq:s_pf_1}, \eqref{eq:s_pf_2} represent the case where the two nodes $i$ and $j$ belong to different load blocks separated by a switched line with status variable $s_{mn}$.
If the load block is energized $z_m=1$ or the switch is closed $s_{mn}=1$, the corresponding pair of constraints form an equality that represents the voltage drop. If the load block is de-energized $z_m=0$ or the switch is open $s_{mn}=0$, the relationship is relaxed using the Big-M method. 
In all of these equations, $p_{lik}$ and $q_{lik}$ refer to the 'sending side' power flows along line $l$. Because we are assuming lossless lines, we have that $p_{lik} = -p_{lki}$ and $q_{lik} = -q_{lki}$.
The voltage magnitudes are kept within bounds by enforcing the following constraint on the squared voltage magnitudes $w_i$, 
\begin{equation}
    (v_i^{min})^2 \leq w_i \leq (v_i^{max})^2~.
\end{equation}

\subsubsection{Limits on Power Sources and Power Flows}
In our model, we assume that power can come either from a substation (i.e., supplied from the bulk grid) or from distributed generation. The amount of power from each source is enforced by the following equations,
\begin{align}
p_{gi}^{min}z_m &\leq p_{gi} \leq p_{gi}^{max}z_m \quad \forall i\in\mathcal{N}, \text{  with } m=NB_{i},\\
q_{gi}^{min}z_m &\leq q_{gi} \leq q_{gi}^{max}z_m \quad \forall i\in\mathcal{N}, \text{  with } m=NB_{i},
\end{align}
where $z_m$ is the energization variable of the bus where the generator is connected and the limits on active and reactive power are given by $p_g^{min}, p_g^{max}$ and $q_g^{min}, q_g^{max}$, respectively. If generator is a substation, we set $p_g^{min}=q_g^{min}=-\infty$ and $p_g^{max}=q_g^{max}=\infty$. 
Further, we limit the amount of power flowing from node $i$ to node $j$ by 
\begin{align}
-p_l^{max}z_{b} &\leq p_{lmn} \leq  p_l^{max}z_{b} \\
-q_l^{max}z_b &\leq q_{lmn} \leq q_l^{max}z_b~\\
&\quad \forall lmn \in\mathcal{L}, \text{ with } NB_m=NB_n=b &\nonumber\\
-p_l^{max}s_{lmn} &\leq p_{lmn} \leq  p_l^{max}s_{lmn}\qquad \forall~{lmn} \in \mathcal{S}\\
-q_l^{max}s_{lmn} &\leq q_{lmn} \leq q_l^{max}s_{lmn}\qquad \forall~{lmn} \in \mathcal{S}
\end{align}
\end{subequations}
where the two first equations represent power flows on lines that are internal to a load block, while the two last represent power flows on switched lines that connect load blocks. 

\subsubsection{Switching Constraints}
We include the switching constraints \eqref{eq:switching} to ensure that energized and de-energized parts of the network are appropriately isolated from one another.

\subsubsection{Radiality Constraints}
Finally, we need to include radiality constraints to ensure that the system operates in a radial topology. There are multiple ways of incorporating radiality constraints, as discussed in detail in the next section.

\section{Efficient Representation of Radiality Constraints in Systems with De-Energization}

In this section, we review two existing methods to represent radiality constraints in distribution grid optimization and propose new versions for each of them. 

\subsection{Existing Parent-Child Representation}
The parent-child radiality constraints \cite{arif2018dynamic, jabr2012minimum} form a (radial) tree structure by ensuring that 
each node in the grid has a single parent, except the substation node which acts as a root node. Parent-child relationships are then be associated with network topology, ensuring radiality. This method can be extended to ensure radiality in islanded networks or partially de-energized networks by use of a \emph{virtual source} proposed in \cite{arif2018dynamic}, which acts as a universal root node (and is a parent to the original substation node). 
We review the equations and variables needed for the representation of parent child relationships.

\subsubsection{Virtual source}
If a distribution system is split into multiple islands, it is necessary to introduce a virtual source node $\mathcal{V}$ in order to ensure radiality in each island. This virtual source node $\mathcal{V}$ is assumed to have connections (i.e., virtual lines) to all other nodes in the system. We denote the virtual line $l$ connecting the virtual source $\mathcal{V}$ to a given node $n$ as $l\mathcal{V}n$

\subsubsection{Parent-child variables}
Parent-child relationships are established between adjacent nodes $m$ and $n$ via binary $\beta$ variables. $\beta_{lmn} = 1$ means that node $m$ is the parent of node $n$, while $\beta_{lnm} = 1$ indicates that node $n$ is the parent of node $m$. If $\beta_{lmn} = \beta_{lnm} = 0$, there is no connection between the two nodes. 

\subsubsection{Parent-child constraints}
We can now define a set of constraints that describe the interconnections between nodes. First, we introduce equations for the virtual node, 
\begin{subequations}
\label{original_pc}
\begin{align}
    \beta_{lm\mathcal{V}} &= 0  &  \forall m \in \mathcal{N},  \label{eq:virt_1} \\
    \beta_{l\mathcal{V}n} &= 1  &  \forall n \in \mathcal{N}_S~, \label{eq:virt_2}
\end{align}
where \eqref{eq:virt_1} ensure that the virtual node is the root node (i.e., has no parents among the other $m$ nodes in the grid) and \eqref{eq:virt_2} 
ensures that the virtual node is the parent for all substations in the grid, represented by the set of substation nodes $\mathcal{N}_S$. 
The remaining parent-child constraints are given by 
\begin{align}
    \sum_{l \in \mathcal{L}_n^- \cup  l\mathcal{V}n}{\beta_{lmn}} &=z_n  &  \forall n &\in \mathcal{N}~, \label{eq:beta1} \\
    \beta_{lmn} + \beta_{lnm} &= 1  &  \forall l &\in \mathcal{L}\backslash \mathcal{S}~, \label{eq:beta2}\\
    \beta_{lmn} + \beta_{lnm} &= s_{lmn}  &  \forall l &\in \mathcal{S}~. \label{eq:beta3}
\end{align}
Here, \eqref{eq:beta1} ensures that each node $n$ has either a single parent (if energized), or no parents (if not). This is done through a sum over all real lines for which $n$ is the 'to' node (represented by the set $\mathcal{L}_n^-)$, in addition to the virtual line connecting the virtual source to $n$ (represented by $l\mathcal{V}n$).
For two nodes $m$ and $m$ connected by a non-switched line $l\in\mathcal{L}\backslash \mathcal{S}$, \eqref{eq:beta2} requires that the parent-child relationship is mutually exclusive, i.e. node $m$ is the parent of node $n$ or node $n$ is the parent of node $m$.   
For two nodes $m$ and $n$ connected by a switched line $l\in\mathcal{S}$, \eqref{eq:beta3} requires that the parent-child relationship between nodes $m$ and $n$ is either mutually exclusive if the switch is closed $s=1$, or that there is no relationship if $s=0$. 



\subsubsection{Virtual flow}
In a system where all nodes are energized by a single substation, the need to transport power from the substation to individual nodes (in combination with the parent-child relationships) will ensure that there is a radial connection between the substation and each node in the network. However, when parts of the network are islanded (e.g., served through distributed generation), the power flow alone no longer serves this purpose. Instead, we introduce a virtual flow based on  \cite{arif2018dynamic} which flows from a virtual source to all energized nodes in the network. 

We denote the virtual flow on line $l$ from node $m$ to node $n$ by $f_{lmn}$ and assign each node $n$ in the system a virtual demand of $d_n=1$. The fictitious line $l\mathcal{V}n$ is introduced to represent virtual flow directly from the virtual source $\mathcal{V}$ to a node $n$. We then enforce the following virtual flow constraints,
\begin{align}
    f_{l\mathcal{V}n}+\sum_{l\in \mathcal{L}_n^-}\!\!\!f_{lmn}  +\! \sum_{l\in\mathcal{L}_n^+}\!{f_{lnm}}  &= 1~ &&\forall n\in\mathcal{N}~, \label{eq:vflow1_na}\\
    0\leq  f_{l\mathcal{V}n} &\leq |\mathcal{N}|\beta_{l\mathcal{V}n} &&\forall l\in\mathcal{L}~, \label{eq:vflow2_na} \\
     -|\mathcal{N}|\leq f_{lmn} &\leq |\mathcal{N}| &&\forall l\in\mathcal{\mathcal{L}\backslash \mathcal{S}}~. \label{eq:vflow3_na}\\
    -|\mathcal{N}|s_{lmn} \leq f_{lmn} &\leq |\mathcal{N}|s_{lmn} &&\forall l\in\mathcal{S}~. \label{eq:vflow4_na}
\end{align}
\end{subequations}
Conservation of virtual flow at each node is expressed by \eqref{eq:vflow1_na}, where the sums over the lines connected to node $n$ are split based on whether $n$ is the 'from' node (for lines $l \in \mathcal{L}_n^+$) or the 'to' node (for lines $l \in \mathcal{L}_n^-$). 
Eq. \eqref{eq:vflow2_na} establishes that virtual flow from $\mathcal{V}$ can only flow to nodes $n$ that are children of $\mathcal{V}$. If $\beta_{l\mathcal{V}n}=0$ (i.e., node $n$ is not a child of $\mathcal{V}$), then $f_{l\mathcal{V}n}=0$. If $\beta_{l\mathcal{V}n}=1$, then the flow can take on any value up to the maximum possible flow, which is equal to the total virtual demand $\sum_{n\in\mathcal{N}}d_n=|\mathcal{N}|$. Eq. \eqref{eq:vflow3_na} sets virtual flow limits to this same maximum on all non-switched lines. \eqref{eq:vflow4_na} does the same for switched lines, and also ensures that there is no virtual flow on lines with open switches ($s_{lmn}=0$).
Since each node can only have one parent and the virtual flow must both reach all nodes and respect borders between system islands, this set of constraint ensures that there is a connection (and corresponding parent-child relationship) between the virtual source and a \emph{single} node within each islanded group of nodes which acts as the root node of that island. With these constraints, a radial structure is then guaranteed in all parts of the network.

\subsection{Proposed Abstracted Parent-Child Formulation}
The proposed, equivalent formulation of the radiality constraints rely on the use of the network abstraction defined in Section \ref{sec:abstraction}.
With this abstraction, we can formulate the radiality constraints with a simplified set of parent-child constraints
\begin{subequations}
\label{abstracted_pc}
\begin{align}    
    \beta_{lm\mathcal{V}} &= 0    &&\forall m \in \mathcal{B}, \label{eq:virt_1_ab} \\
    \beta_{l\mathcal{V}n} &= 1  &&  \forall n \in \mathcal{B}_S~. \label{eq:virt_2_ab} \\
    \sum_{l \in \mathcal{L}_n^- \cup  l\mathcal{V}n}\!\!\!{\beta_{lmn}} &=z_n  &&  \forall n \in \mathcal{B}~, \label{eq:beta1_ab} \\
    \beta_{lmn} + \beta_{lnm} &= s_{lmn}  &&  \forall l \in \mathcal{S}~, \label{eq:beta3_ab}
\end{align}
and a simplified set of virtual flow constraints, 
\begin{align}
    f_{l\mathcal{V}n}+\sum_{l\in \mathcal{L}_n^-}\!\!\!f_{lmn}  +\! \sum_{l\in\mathcal{L}_n^+}\!{f_{lnm}}  &= 1~ &&\forall n\in\mathcal{B}~, \label{eq:vflow1_ab}\\
    0\leq  f_{l\mathcal{V}n} &\leq |\mathcal{B}|\beta_{l\mathcal{V}n} &&\forall l\in\mathcal{S}~, \label{eq:vflow2_ab} \\
    -|\mathcal{B}|s_{lmn} \leq f_{lmn} &\leq |\mathcal{B}|s_{lmn} &&\forall l\in\mathcal{S}~. \label{eq:vflow3_ab}
\end{align}
\end{subequations}
Note that these constraints are similar to the constraints for the standard network, except that we consider load blocks $\mathcal{B}$ instead of individual nodes $\mathcal{N}$ and the abstracted network omits any consideration of non-switched lines $\mathcal{L}\backslash\mathcal{S}$.

\subsection{Existing Loop-Based Formulation}
Another popular way of ensuring radiality is to identify the set of loops that exists in the network if all switches are closed, denoted by the set $\mathcal{O}$, and then implement constraints which ensure that at least one switch around the loop will be open. 
This algorithm proceeds in two steps.

First, we identify the set $\mathcal{O}$ containing all the loops in the system. This can be done by utilizing the loop-enumeration algorithm in \cite{loop_enum_algo}, which was also applied in \cite{arif_thesis}. 

Second, we use the knowledge regarding the set of possible loops to implement constraints in our problem. 
Once we have identified the set of loops $\mathcal{O}$, we run a post-processing to identify the set of switches $\mathcal{S}_i$ associated with each loop $i\in\mathcal{O}$. Using this information, we enforce that
\begin{equation}
\label{eq:loops}
    \sum_{l \in \mathcal{S}_i} s_{lmn} \leq |\mathcal{S}_i| - 1\quad \forall i \in \mathcal{O}~,
\end{equation}
which guarantees that at least one switch in each loop must be open, thus preventing the formation of actual loops.

\subsection{Proposed Constraint Generation Algorithm for Loop-based Formulation}


While the loop constraints themselves are conceptually very simple, the difficulty of the enumerating all possible loops grows extremely quickly with network size and number of switches. To mitigate this issue, we propose an iterative algorithm based on constraint generation for systems with a larger number of loops. 
This iterative algorithm adds a third term to the objective function specified by (\ref{eq:objective}) that acts as a penalty on closed switches, i.e., 
\begin{align} 
min \sum_{b \in \mathcal{B}} R_b z_b(1-\alpha)
- \sum_{b \in \mathcal{B}}D_b z_b \alpha + \gamma \sum_{l\in\mathcal{S}} s_{lmn}
\label{eq:objective_2}
\end{align}
By choosing a penalty value $\gamma$ that is large enough to encourage switches to be open, yet small enough to allow switches to close when necessary to serve load, it is possible to discourage solutions with loops without interfering with the main aspects of the problem that is solved. It should be noted that $\gamma$ must be empirically determined and may vary widely between different systems.
We use this observation to devise an iterative algorithm for the loop-based radiality constraints. 



\noindent\textbf{\emph{Step 0:}} Initialize the algorithm with an iteration count $k=1$ and an empty set of loops $\mathcal{O}^{(1)}=\{\}$.

\noindent\textbf{\emph{Step 1:}}
Solve an adapted version of the optimization problem
with objective \eqref{eq:objective_2}
and relaxed radiality constraints
\begin{equation}
    \sum_{l \in \mathcal{S}_i} s_{lmn} \leq |\mathcal{S}_i| - 1\quad \forall i \in \mathcal{O}^{(k)}~,
    \label{eq:loops_limited}
\end{equation}

\noindent\textbf{\emph{Step 2:}} Given the switching status $s^{(k)}$ obtained with the solution from Step 1, generate the system topology and run the loop-enumeration algorithm to identify the set of observed loops $\hat{\mathcal{O}}^{(k)}$. If there are no loops in the solution $|\hat{\mathcal{O}}|^{(k)}=0$, terminate the algorithm. If $|\hat{\mathcal{O}}|^{(k)}\geq 1$, add the identified loops to the considered set of loops in the next iteration
$\mathcal{O}^{(k+1)}=\mathcal{O}^{(k)}\cup \hat{\mathcal{O}}^{(k)} $, increase the iteration count to $k=k+1$ and return to Step 1.

\section{Comparison of Problem Variations}
In this section, we compare the different methods for representing radiality constraints.
We implement the following optimization models:\\
\textbf{I) Original Parent-Child (Original P-C)} includes the objective \eqref{eq:objective}, grid constraints \eqref{opf-formulation} and switching constraints \eqref{eq:switching}, as well as the original (non-abstracted) parent-child radiality constraints \eqref{original_pc}. \\
\textbf{II) Abstracted Parent-Child (Abstracted P-C)} includes \eqref{eq:objective}, \eqref{opf-formulation} and  \eqref{eq:switching}, as well as the abstracted parent-child radiality constraints \eqref{abstracted_pc}. \\
\textbf{III) Naive Loop-Based (Naive L-B)} includes \eqref{eq:objective}, \eqref{opf-formulation} and  \eqref{eq:switching}, as well as the loop-based constraint \eqref{eq:loops} for all possible loops in the system. \\
\textbf{IV) Iterative Loop-Based (Iterative L-B)} includes the objective with a penalty on closed switches \eqref{eq:objective_2}, \eqref{opf-formulation} and  \eqref{eq:switching}, as well as a limited set of the loop-based constraints \eqref{eq:loops_limited} that is built iteratively. We set the penalty parameter $\gamma=10^{-6}$.\\[+2pt]
%
%
All OPF formulations were written in Julia using the JuMP mathematical optimization package \cite{JuMP}, and solved with the Gurobi solver \cite{gurobi}. All Gurobi settings were left at the default.
The computing platform was a Macbook Pro containing a 3.1 GHz Dual-Core Intel Core i7 and 16GB of DDR3 RAM.

\subsection{Test Case} \label{sec:testcase}
We create a test case based on the IEEE 123-bus case. The original documentation for the system, was used to add in switched lines missing from the standard case. Both the case and documentation are available at \cite{123_bus_case}. Some of these switches enable internal topology changes within the system, while other indicate connections to other feeders. In order to create a larger test system, these "external" switched lines were used to connect 16 copies of the 123 bus case to each other. Generators were placed at the endpoints of switched lines that were not connected to another copy of the network, in order to mimic an interconnection to an even larger system. We also created smaller versions with 2, 4, and 8 copies 
to enable a discussion of method scalability. The system topologies and associated data can be found in \cite{cases}.

For simplicity, the risk values for the lines $R_l$ are obtained by random sampling as described below. To generate synthetic wildfire risk data, we first sample risk values uniformly between 1 and 10 for each load block in the network. This is done to simulate the often significant variability in risk between adjacent regions (e.g. zero risk for undergrounded lines in an urban area adjacent to overhead lines through a nearby wooded area). Risks for individual lines within load blocks are then sampled from normal distributions centered around the block risk value, with a variance of 0.25. Switches were considered to be short line segments with zero risk. We note that although we choose to work with randomly generated risk values, those values could easily be replaced by more realistic data obtained from historical records of wildfire risk, as described in e.g. \cite{taylor2021_wildfire_risk}.

The OPF formulations were compared 
with trade-off values $\alpha$ ranging between 0.0 and 1.0 in increments of 0.1. For each value of $\alpha$, 10 seed values were varied to produce 10 different problem instances with varying network risk profiles. 

A sample plot for a solution of the 16-copy case, created using the PowerPlots package for Julia \cite{powerplots}, is shown below.
\begin{figure}
    \centering
    \includegraphics[scale = 0.58]{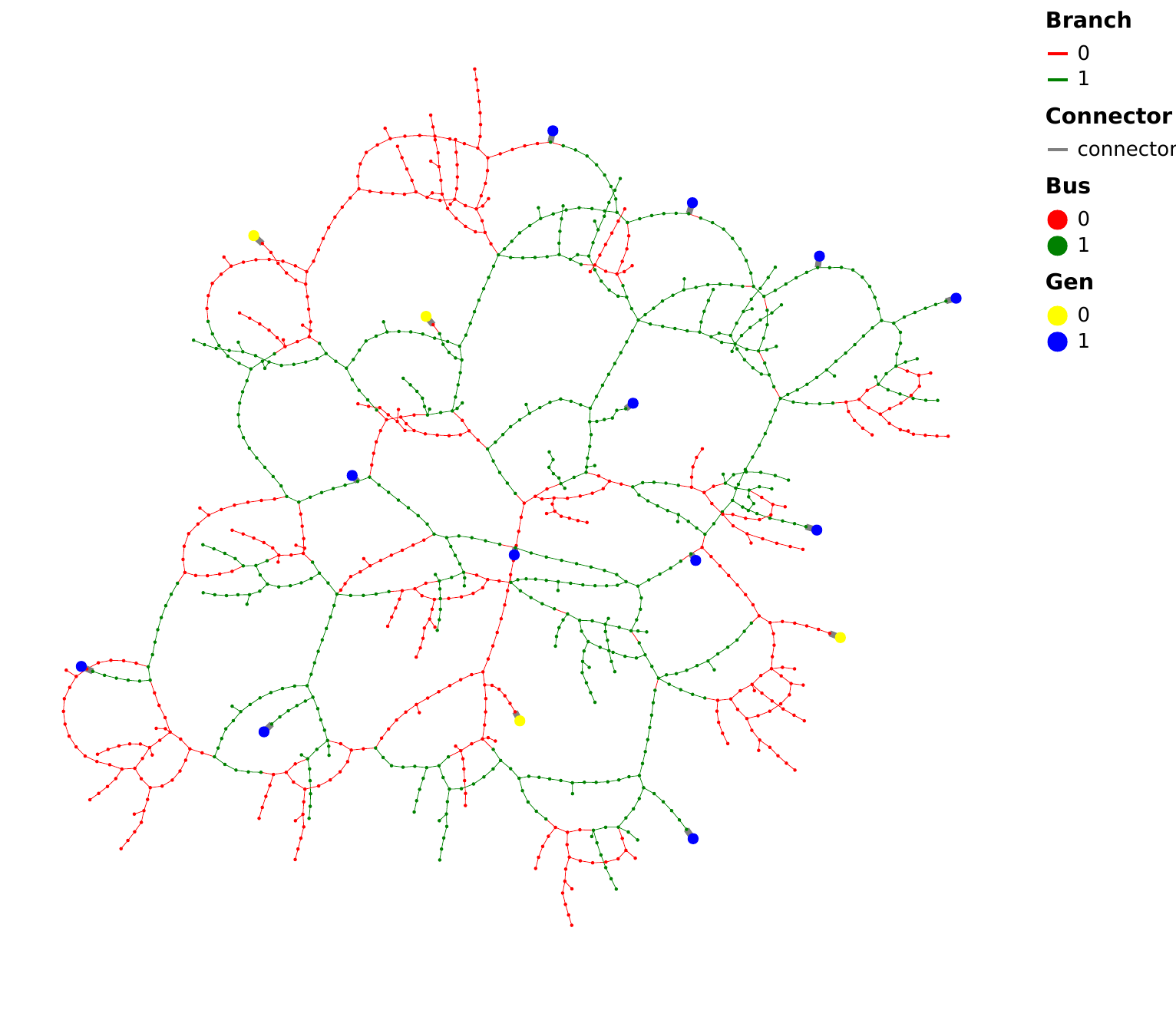}
    \vspace{-30pt}
    \caption{Power shut-off solution for the 16 copy case with the second seed for $\alpha=0.5$. The nodes and lines indicated in red are de-energized parts of the network, while the green nodes and lines are still operating. Substation nodes are indicated in blue (if connected) or yellow (if not). }
    \label{fig:my_label}
\end{figure}


\subsection{Comparison of Radiality Constraint Formulations}
We first compare important characteristics of the radiality constraints,
including the problem size and number of loops.

\subsubsection{Problem size with abstracted and non-abstracted parent-child constraints}
The main motivation for applying parent-child constraints to the abstracted network was to increase efficiency through a reduction in the number of necessary binary variables. Thus the primary point of comparison between the abstracted and original parent-child formulations is the number of binary variables as a function of network size. The Original P-C formulation requires $(3|\mathcal{N}|+2|\mathcal{L}|+|\mathcal{S}|+|\mathcal{N}_S|)$ constraints,
$(2|\mathcal{L}|+|\mathcal{N}|)$ binary variables, and $(2|\mathcal{L}|)$ continuous variables. The Abstracted P-C requires $((3|\mathcal{B}|+3|\mathcal{S}|+|\mathcal{N}_S|))$ constraints, $(2|\mathcal{S}|+|\mathcal{B}|)$ binary variables and $(2|\mathcal{S}|)$ continuous variables. 

To make the comparison concrete, we provide the above numbers for each formulation and for 1, 2, 4, 8 and 16 connected copies of the IEEE 123-bus case in Table \ref{tab:variable_comparison}. We observe that the radiality constraints in the Original P-C requires between 6 and 10 times more constraints and variables than the Abstract P-C. The absolute difference in number of variables and constraints is particularly striking for the largest 16 copy system. The difference arises because only a small fraction of the total lines $\mathcal{L}$ have switches $\mathcal{S}$ and the set of buses $\mathcal{N}$ is necessarily larger than the set of load blocks $\mathcal{B}$ for a given network.


\begin{table}
\centering
\caption{Variable Comparison Between Original and Abstracted Parent-Child Methods}
\begin{tabular}{lccccc}
\hline
 & \textbf{Case 1} & \textbf{Case 2} & \textbf{Case 4} & \textbf{Case 8} & \textbf{Case 16} \\ \hline
 \textbf{Buses $\mathcal{N}$} & 71 & 135 & 268 & 536 & 1072 \\ 
\textbf{Lines $\mathcal{L}$} & 72 & 139 & 277 & 557 & 1118 \\ 
\textbf{Switches $\mathcal{S}$} & 11 & 16 & 33 & 69 & 142 \\ 
\textbf{Load Blocks $\mathcal{B}$} & 9 & 11 & 24 & 48 & 96 \\[+3pt] \hline

\multicolumn{6}{l}{{ \textbf{Original Parent-Child}}} \\ \hline
 \textbf{Constraints} & 369 & 700 & 1392 & 2792 & 5595 \\ 
\textbf{Binary } & 215 & 413 & 822 & 1650 & 3308 \\ 
\textbf{Continuous } & 144 & 278 & 554 & 1114 & 2236 \\[+3pt] \hline

\multicolumn{6}{l}{{ \textbf{Abstracted Parent-Child}}} \\ \hline
 \textbf{Constraints} & 61  & 82 & 172 & 352 & 715 \\
\textbf{Binary } & 31  & 43 & 90 & 186 & 380 \\ 
\textbf{Continuous } & 22  & 32 & 66 & 138 & 284 \\ \hline
\end{tabular}
\label{tab:variable_comparison}
\end{table}

\subsubsection{Number of Loops}
The Naive L-B formulation requires that all potential loops in the system are enumerated, and that constraints of the form \eqref{eq:loops} are enforced for each one. 
As the number of loops in a system relies on its topology, in addition to the number of nodes, edges and switches, a general formula for loop count is impractical. Instead, Table \ref{tab:loop_enumeration} lays out the number of loops found in the 1, 2, 4, 8 and 16 copy cases, as well as the computational time required to do the enumeration. 
For the original system, we only identify 3 loops and the total computation time is less than a second. For case 4, the numbers are 180 loops and 2.39 s, while case 8 identifies 151 632 loops and requires close to 11 000 s. Enumeration for case 16 could not be completed in a reasonable amount of time (designated to be 12 hours, or 43,200 seconds), but within this time, the enumeration algorithm identified close to 600 000 loops. 
These results indicate that the number of loops, and the time required to compute them, increased dramatically with the size of the network, which is not unexpected. 

\begin{table}
\caption{Loop Enumeration in the Naive Loop-Based Approach}
\begin{tabular}{l|ccccc}
\hline
 & \textbf{Case 1} & \textbf{Case 2} & \textbf{Case 4} & \textbf{Case 8} & \textbf{Case 16} \\ \hline
\textbf{Loops} & 3 & 18 & 180 & 151632 & 599,962+ \\ 
\textbf{Time (s)} & 0.38 & 0.45 & 2.39 & 10992.85 & 43,200+ \\ \hline

\end{tabular}
\label{tab:loop_enumeration}
\end{table}

\subsection{Numerical Comparison}
We next solve the distribution grid optimization models making use of three different radiality constraint formulations, the original P-C, the abstracted P-C, and the iterative L-B,  were solved for the 16-copy case. 
The naive L-B formulation is not included in this comparison due to the prohibitively long time required for the loop enumeration/constraint generation step.
We solve the problem for 10 different seed values and the full range of alpha values from 0 to 1 in increments of 0.1, giving us a total of 100 problem instances. We compare only the solution times, as all methods converged to the same optimal solution in each instance.
The results are summarized in figure \ref{fig:main_result_graph} below, which plots OPF solution time against the risk-load trade-off value $\alpha$. Table \ref{tab:main_result_table} gives 
median, minimum and maximum solve-time (across the 10 seed values tested) for each method and selected values of $\alpha$.

\begin{table}
\centering
\caption{Comparison of Solution Times for Selected $\alpha$ Values}
\begin{tabular}{c|l|ccccc}
\hline \hline
\multicolumn{2}{l}{\textbf{Trade-Off Value} $\boldsymbol{\alpha}$}  & \textbf{0.10} & \textbf{0.30} & \textbf{0.50} & \textbf{0.70} & \textbf{0.90} \\ \hline \hline 
\multirow{3}{*}{\textbf{\begin{tabular}[c]{@{}c@{}}Original  \\ P-C\end{tabular}}} & Median [s] & 1.11 & 3.06 & 18.01 & 173.39 & 198.53 \\ 
 & Min [s] & 0.91 & 1.71 & 6.62 & 84.53 & 145.91 \\ 
 & Max [s] & 1.45 & 3.91 & 50.72 & 383.92 & 4111.75 \\ \hline
\multirow{3}{*}{\textbf{\begin{tabular}[c]{@{}c@{}}Abstracted \\ P-C\end{tabular}}} & Median [s] & 0.53 & 1.67 & 6.38 & 9.42 & 9.24 \\ 
 & Min [s] & 0.50 & 1.08 & 1.10 & 3.89 & 4.98 \\ 
 & Max [s] & 0.95 & 3.37 & 32.62 & 15.71 & 18.14 \\ \hline
\multirow{3}{*}{\textbf{\begin{tabular}[c]{@{}c@{}}Iterative \\ L-B\end{tabular}}} & Median [s] & 0.66 & 1.94 & 6.37 & 12.12 & 7.54 \\ 
 & Min [s] & 0.54 & 1.07 & 1.34 & 2.89 & 2.82 \\ 
 & Max [s] & 1.84 & 6.29 & 51.26 & 36.69 & 43.53 \\ \hline \hline
\end{tabular}
\label{tab:main_result_table}
\end{table}

\begin{figure}
    \centering
    \includegraphics[scale=0.68]{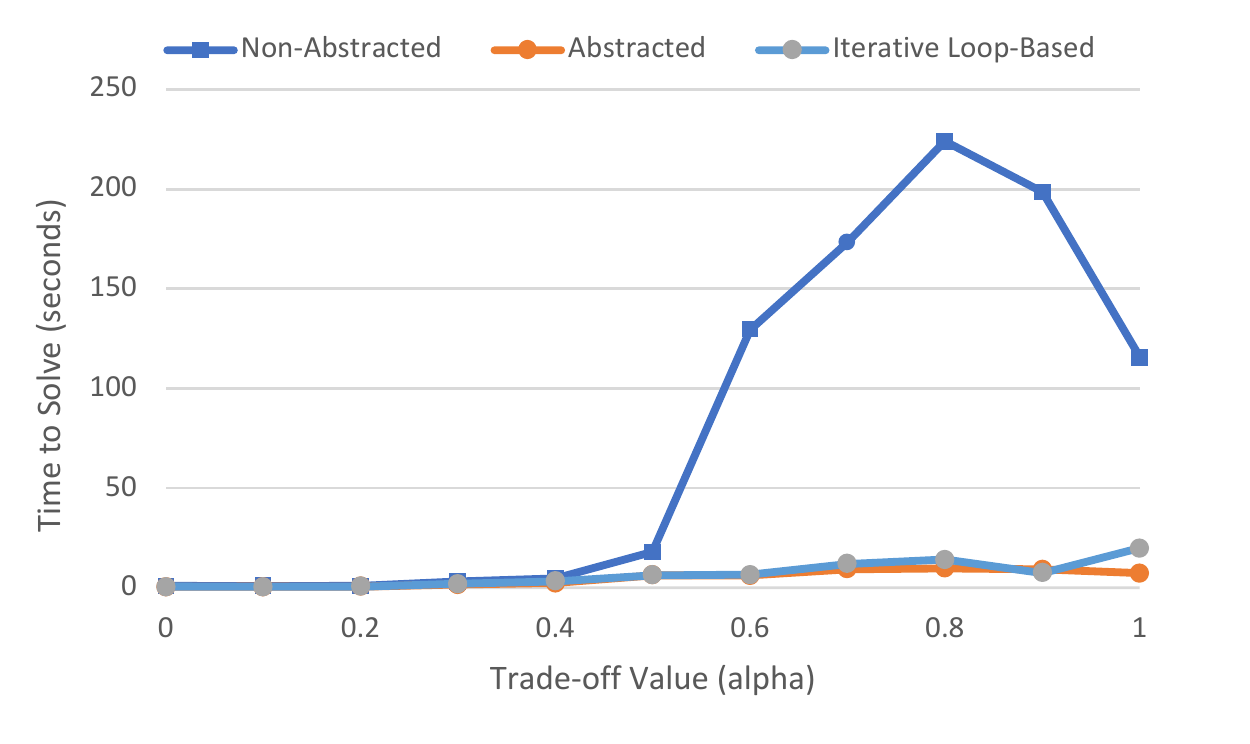}
    \vspace{-5pt}
    \caption{Comparison of Solution Times for the Original P-C (dark blue), the Abstracted P-C  (orange) and the Iterative L-B  (light blue) for $\alpha$ from 0 to 1. Plot shows the median solution times across 10 different risk profiles.}
       \vspace{-8pt}
    \label{fig:main_result_graph}
\end{figure}

The optimization model containing the abstracted P-C radiality constraints consistently shows significantly lower computational time compared with original P-C model, especially for $\alpha$ values between 0.5 and 0.9. In this range, the abstracted P-C were on average 20 times faster. For $\alpha=0.9$, the computation times ranged from 4.98 to 19.18 s for the abstracted P-C, while the original P-C required from 145.91 to 4111.75 s. 

The iterative L-B formulation also  performed quite well, yielding solution times nearly as fast as those of the abstracted parent-child method. It should be noted that these times include both any necessary loop enumeration, and time taken by the solver after loop constraint generation is complete (across all iterations of the algorithm). Contrasting this performance with the 12+ hours necessary for the naive L-B formulation to enumerate all possible loops suggests that the iterative method must only enumerate a small fraction of these loops. 
While the number of enumerated loops vary between the different seed and $\alpha$ values, the algorithm only has to enumerate between 0 and 135 loops. We note that the iterative L-B formulation did require some careful tuning penalty factor $\gamma$. Setting $\gamma=10^{-6}$ was enough to discourage excessively meshed intermediate solutions in the iteration process, while not causing changes to the optimal solution. 

\subsection{Analyzing Optimal Shut-Off Solutions}
We finally analyze solution sensitivity to two main problem parameters, the trade-off value $\alpha$ and the risk profile.

\begin{figure}
    \centering
    \includegraphics[scale=0.4]{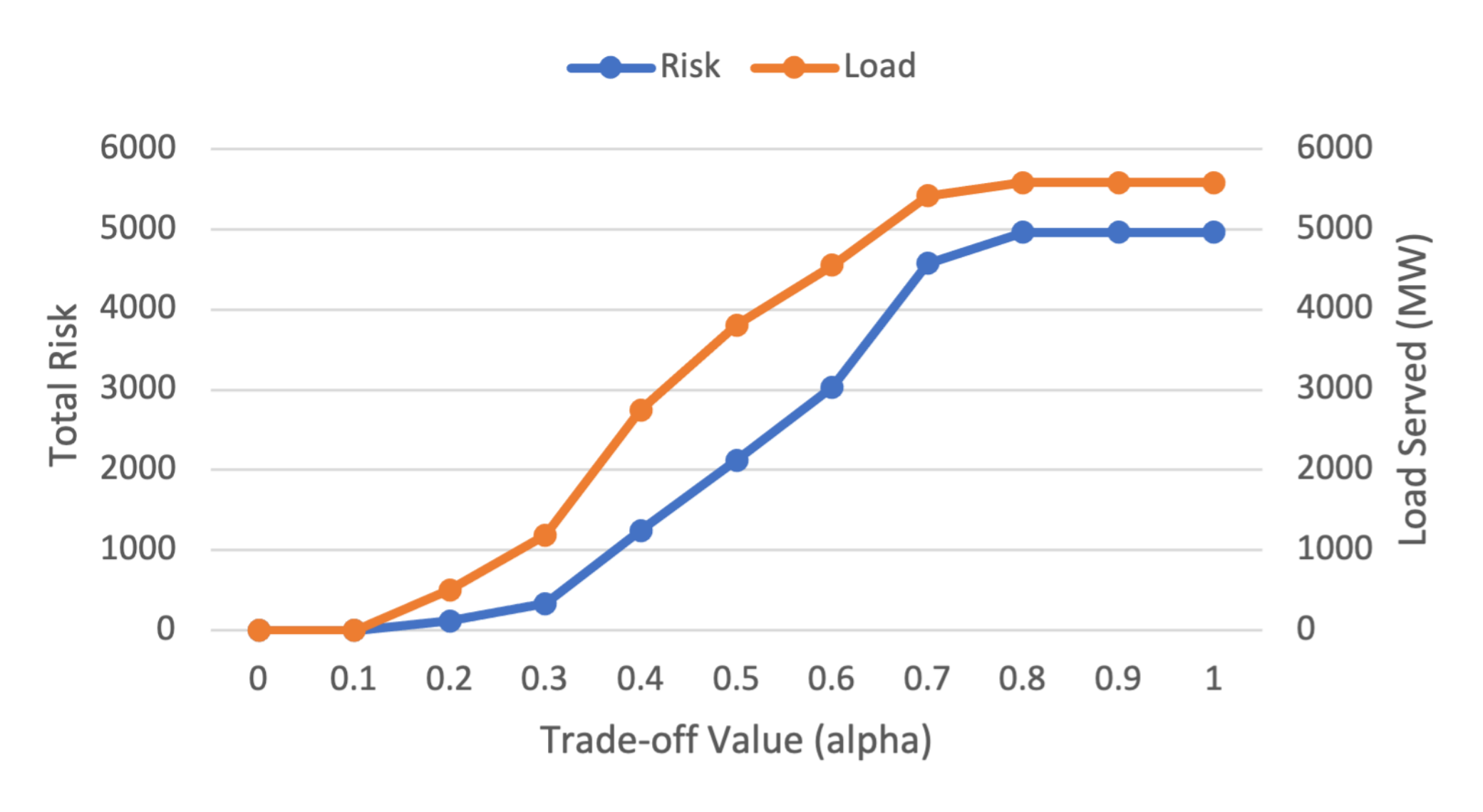}   
    \vspace{-5pt}
    \caption{Wildfire risk and load served for different values of $\alpha$.}
       \vspace{-8pt}
    \label{fig:SolVarAlpha}
\end{figure}

First, we analyze how the solution changes for different values of $\alpha$. We solve the Abstract P-C model for case 16 with a seed value of 1.
In Figure \ref{fig:SolVarAlpha}, the system risk profile is held constant and aggregate system risk and load served are plotted against values of $\alpha$ from 0 to 1. As $\alpha$ increases, more and more importance is placed on load served in place of risk minimization, explaining the general rise in both factors. Notably, the  gap between load and risk increases from $\alpha = 0$ to $\alpha = 0.5$, before decreasing for $\alpha > 0.5$. This characteristic is explained by the fact that, while some increase in risk is necessary to serve more load, the prioritization of risk minimization can depress this behavior. Overall, Figure \ref{fig:SolVarAlpha} provides confirmation that our models behave as expected.

\begin{figure}
    \centering
    \includegraphics[scale=0.4]{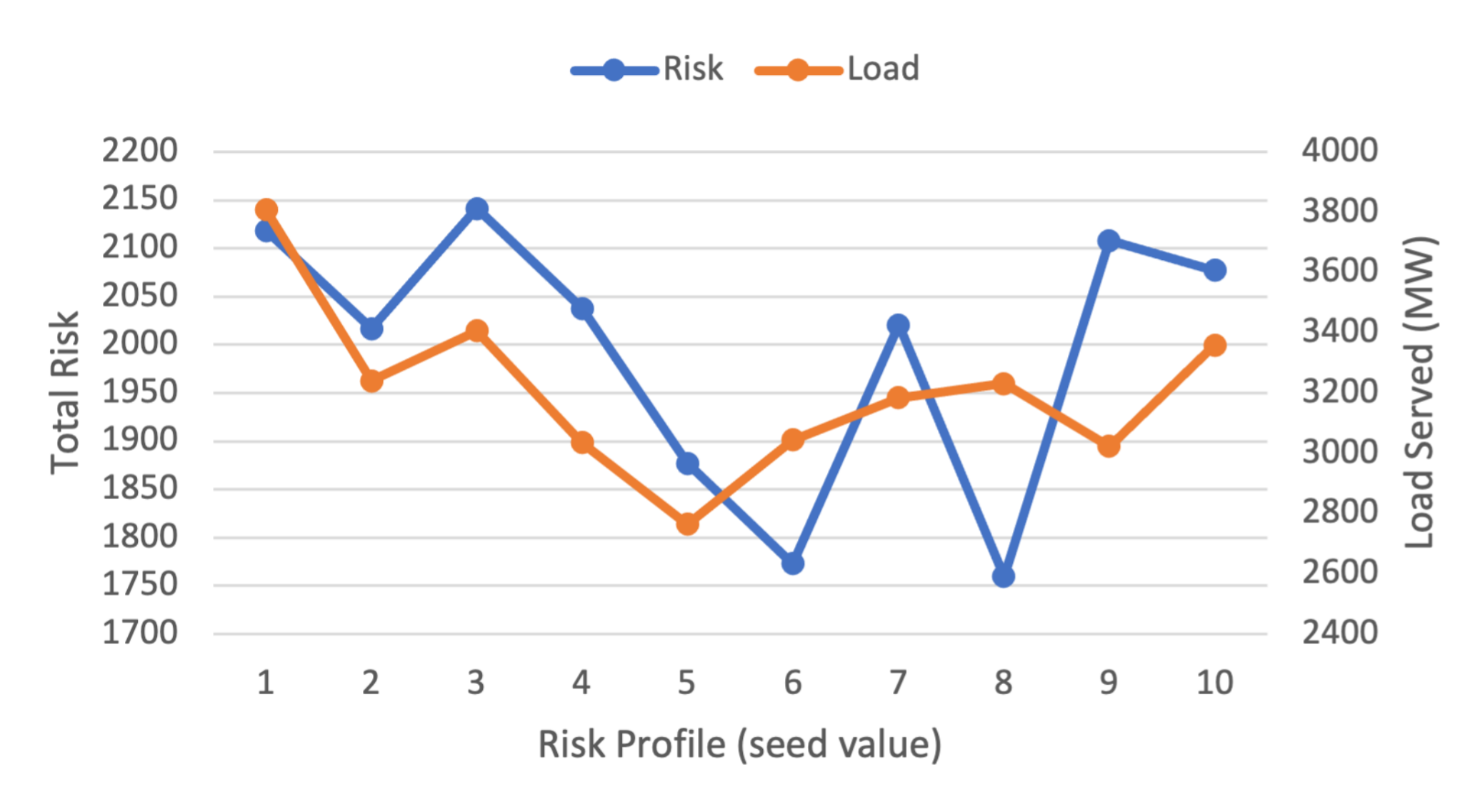}
       \vspace{-5pt}
    \caption{Wildfire risk and load served for different risk profiles.}
    \label{fig:SolVarRisk}
       \vspace{-8pt}
\end{figure}

Next, we vary the risk profile while keeping $\alpha$ constant at $\alpha=0.5$ and varying the risk profile by varying the random seed value. 
Figure \ref{fig:SolVarRisk} shows how risk and load vary as the risk profile vary with the different seeds. We observe that the different risk profiles lead to very different solutions. 
Such sensitivity to risk profile highlights the necessity of up-to-date risk data and fast optimization to obtain the most effective power shut-off solutions. 


\section{Conclusions}
This paper proposed two new methods to represent radiality constraints in distribution system optimization, and benchmarked them against existing approaches by solving an optimal power shut-off problem for wildfire risk mitigation on a medium-sized distribution test case. We find that the proposed methods achieve a speed-up of 10-20 times on challenging instances, while identifying the same optimal solutions as existing methods. 

These promising results provide several opportunities for future work. In particular, we would like to test the methods on even larger systems and consider problems which require simultaneous consideration of multiple time steps. Further, we would like to investigate other problem variants and the inclusion of more complex power flow formulations, including three-phase unbalanced models and models which include a more realistic treatment of losses.

\bibliographystyle{IEEEtran}
\bibliography{refs, XX_bib-all}

\end{document}